\pdfoutput=1
\documentclass[lettersize,journal]{IEEEtran}
\usepackage{amsmath,amsfonts}
\usepackage{algorithm}
\usepackage{array}
\usepackage[caption=false,font=normalsize,labelfont=sf,textfont=sf]{subfig}
\usepackage{textcomp}
\usepackage{stfloats}
\usepackage{url}
\usepackage{verbatim}
\usepackage{graphicx}
\usepackage{cite}
\usepackage{booktabs}
\usepackage{tabularx}
\usepackage{multirow}
\usepackage{enumitem}

\usepackage{amsthm}

\usepackage{tabularx,booktabs,array,makecell}
\usepackage{pifont}

\newcolumntype{Y}{>{\centering\arraybackslash}X}

\usepackage{siunitx}
\usepackage{threeparttable}
\usepackage[normalem]{ulem}
\useunder{\uline}{\ul}{}

\DeclareSIUnit{\belmilliwatt}{Bm}
\DeclareSIUnit{\dBm}{\deci\belmilliwatt}
\DeclareSIUnit{\sample}{Sa}

\usepackage{orcidlink}
\hypersetup{
	colorlinks=true,
	linkcolor=black,
	citecolor=black,
	urlcolor=black
}

\usepackage{algpseudocode}
\usepackage{hyperref}

\begin{document}

\title{Simulation-Assisted Radio Map Reconstruction with Sparse Field Measurements}

\author{Lang~Ming\orcidlink{0009-0006-4563-9889},
	~Longhui~Wang\orcidlink{0000-0002-7162-0229},
	~Yazhou~Sun\orcidlink{0009-0001-8633-2334},
	~Xi~Chen\orcidlink{0000-0003-3671-6425},~\IEEEmembership{Senior~Member,~IEEE},
	~Jian~Wang\orcidlink{0000-0001-7683-6937},~\IEEEmembership{Senior~Member,~IEEE},
	~Yu~Wang\orcidlink{0000-0001-6108-5157},~\IEEEmembership{Fellow,~IEEE}
\thanks{Manuscript received 6 Nov. 2026. (Corresponding authors: Jian Wang; Longhui Wang.)}
\thanks{Lang Ming, Longhui Wang, Yazhou Sun, Jian Wang and Yu Wang are with the Department of Electronic Engineering, Tsinghua University, Beijing 100084, China (e-mail: langming2002@outlook.com; wanglonghui@mail.tsinghua.edu.cn; sunyz22@mails.tsinghua.edu.cn; jian-wang@tsinghua.edu.cn;  yu-wang@tsinghua.edu.cn).}
\thanks{Xi Chen is with the Beijing National Research Center for Information Science and Technology, Tsinghua University, Beijing 100084, China (e-mail: chenxiee@tsinghua.edu.cn).}
}

\newcommand\submittedtext{%
	\centering\footnotesize This work has been submitted to the IEEE for possible publication. Copyright may be transferred without notice, after which this version may no longer be accessible.}
\newcommand\submittednotice{\begin{tikzpicture}[remember picture,overlay]
		\node[anchor=south,yshift=10pt] at (current page.south) 
		{\parbox{\dimexpr\textwidth-\fboxsep-\fboxrule\relax}{\submittedtext}};
	\end{tikzpicture}%
}

\markboth{Journal of \LaTeX\ Class Files,~Vol.~14, No.~8, August~2021}%
{Shell \MakeLowercase{\textit{et al.}}: A Sample Article Using IEEEtran.cls for IEEE Journals}

\maketitle
\submittednotice
\IEEEdisplaynontitleabstractindextext
\IEEEpeerreviewmaketitle

\begin{abstract}
Field measurements are essential for evaluating radio map reconstruction under real-world conditions, but their acquisition is costly. This letter proposes a simulation-assisted parameterized-background data-weak (SA-PBDW) method that combines propagation simulations with sparse field measurements, which jointly estimates background coefficients and a spatial correction to account for discrepancies between simulated and measured power distributions. Dense field measurements are collected for multiple transmitter positions at an urban site. Evaluations on the collected data and RadioMapSeer demonstrate improved reconstruction accuracy and structural fidelity over benchmark methods at reduced sampling rates. 
\end{abstract}


\section{Introduction}\label{sec:introduction}
Radio mapping is a promising approach to environment-aware planning and resource management in future wireless networks \cite{romero2022rme}. A radio map represents the spatial distribution of a wireless channel metric, such as average received power or channel gain, over a geographical region \cite{levie2021radiounet}. Its practical utility depends on how accurately it characterizes the propagation conditions in the target environment.

A variety of methods have been developed for radio map reconstruction. Spatial interpolation approaches, including inverse distance weighting (IDW)\cite{Denkovski2012idw}, Kriging \cite{sato2017kriging}, and Gaussian process regression (GPR)\cite{zhang2022gprradio}, estimate values at unobserved locations by exploiting spatial proximity or statistical dependence among observations. Structured recovery methods formulate map reconstruction as matrix or tensor completion, exploiting low-rank and spatial smoothness to recover missing entries \cite{zhang2020cbtd}. Neural approaches employ convolutional \cite{levie2021radiounet}, attention-based \cite{sun2026dun}, or generative architectures \cite{wang2024twostage,zhao2025radiodiff3d,wang2025radiodiff} to predict maps from environmental information or reconstruct them from sparse observations. Despite this methodological diversity, most reported evaluations rely heavily on simulated radio maps \cite{Shrestha2023rme}, while field validation remains comparatively limited. Uncertain material parameters and approximations in propagation modeling can produce discrepancies between simulated and measured distributions. Therefore, performance on simulation benchmarks alone provides insufficient evidence of reconstruction fidelity under real-world propagation conditions.

Recent studies have increasingly incorporated field measurements. These studies include reconstruction based on path loss model, evaluated in a controlled-source experiment in open areas \cite{sun2025accuraterme}, and reconstruction guided by simulated power distributions in urban environments \cite{chen2025urban3dradiomap,cheng2027multifidelity}. These non-neural methods offer practical solutions when measured data are insufficient for neural network training. Nevertheless, reconstruction fidelity remains challenging under sparse spatial sampling. Although these methods can recover the overall power distribution and reveal discrepancies between simulated and measured maps, the reconstructions may smooth out or miss local variations observed in field measurements. Denser measurements can improve reconstruction fidelity and provide more reliable spatial references for evaluation, but their acquisition requires substantial time and effort \cite{wang2024twostage}, hindering the construction of large-scale field datasets. Therefore, a critical challenge in field validation and dataset construction is to minimize acquisition requirements while retaining the spatial characteristics of the real-world power distribution.

To address this challenge, we propose simulation-assisted parameterized-background data-weak (SA-PBDW) reconstruction. Within the PBDW framework \cite{maday2015parameterized}, sparse measurements jointly determine a simulation-based background and a regularized spatial update. Dense field measurements provide detailed references for examining differences between simulated and measured distributions and evaluating reconstruction fidelity as the sampling rate decreases. The main contributions are as follows:
\begin{enumerate}
	\item SA-PBDW is proposed as a non-neural method that combines propagation simulations with sparse field measurements through joint estimation of background coefficients and a regularized spatial update.
	\item Dense outdoor power measurements are collected to construct FieldRadioMap, providing spatially resolved references that reveal local variations and discrepancies from corresponding simulations. The dataset is planned for public release\footnote{The dataset is available on Zenodo (DOI: \href{https://doi.org/10.5281/zenodo.23974604}{10.5281/zenodo.23974604}).}.
	\item Evaluations on FieldRadioMap and RadioMapSeer \cite{levie2021radiounet} demonstrate improved reconstruction accuracy and structural fidelity over classical and state-of-the-art methods under sparse sampling.
\end{enumerate}

\textit{Notations}: Throughout this letter, scalars are denoted by regular lowercase or uppercase, while vectors, matrices are denoted by $\boldsymbol{x}$ and $\boldsymbol{X}$, respectively. Superscript $(\cdot)^\top$ denotes transposition. ${X}_{i,j}$ and $\boldsymbol{X}_{i,:}$ denote the $(i,j)$-th entry and the $i$-th row of $\boldsymbol{X}$, respectively. Specially, $\boldsymbol{e}_i$ is the $i$-th standard basis vector of the required dimension. $range(\cdot)$ and $tr(\cdot)$ denote column space and trace, and $\|\cdot\|_2$ is the Euclidean norm. The cardinality of the set $\mathcal{L}$ is denoted by $|\mathcal{L}|$. For $\boldsymbol{K}\succ0$, define $\langle\boldsymbol{p},\boldsymbol{q}\rangle_K=\boldsymbol{p}^\top\boldsymbol{K}^{-1}\boldsymbol{q}$ and $\|\boldsymbol{p}\|_K^2=\langle\boldsymbol{p},\boldsymbol{p}\rangle_K$.

\section{Proposed SA-PBDW Reconstruction}\label{sec:methods}
\begin{figure*}[!t]
	\centering
	\includegraphics[width=\linewidth]{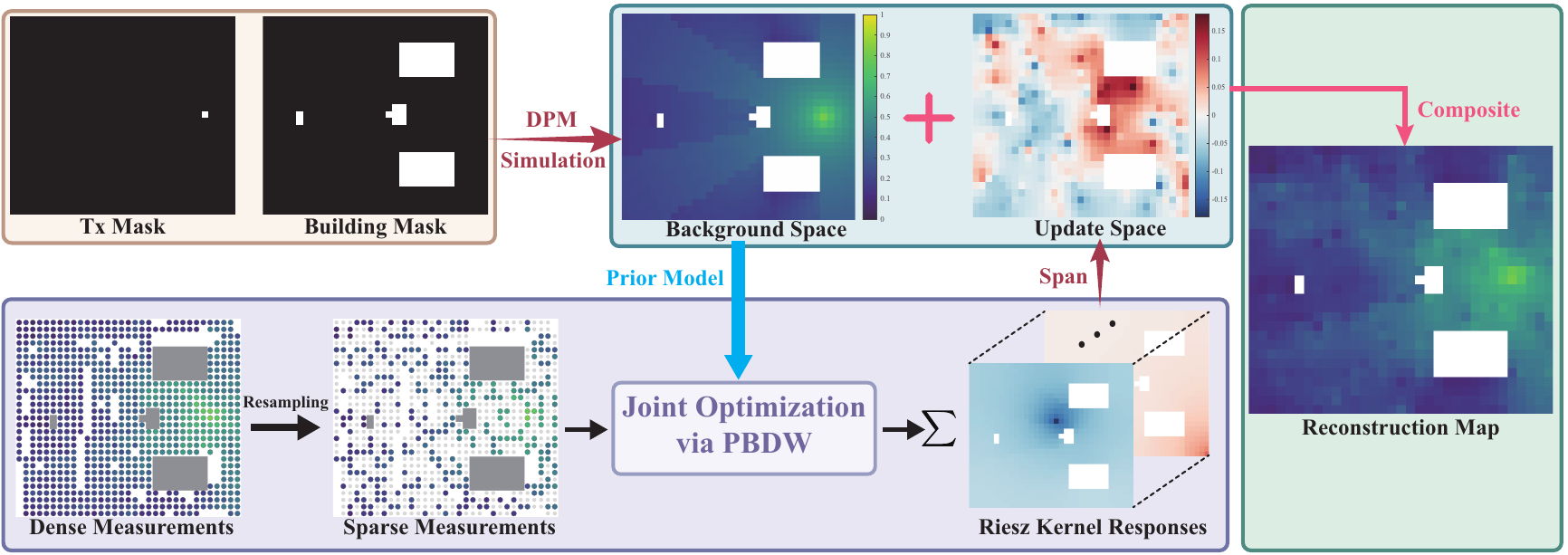}
	\caption{Overall framework of the proposed SA-PBDW method.}\label{fig_pbdw_schema}
\end{figure*}
\subsection{Overall Framework}
Fig.~\ref{fig_pbdw_schema} illustrates the proposed SA-PBDW framework. Given the transmitter position and building layout, the dominant path model (DPM) \cite{wahl2005dominant} generates a low-fidelity (LF) power map. SA-PBDW combines the LF map with sparse high-fidelity (HF) observations from field measurements to reconstruct the target map. Within the PBDW framework, the reconstructed field comprises a simulation-informed background and a correction in a measurement-dependent update space.

For the mathematical formulation, the complete two-dimensional grid contains $N$ non-building cells, indexed in row-major order by $\mathcal{V}=\{1,\ldots,N\}$. Their target power values are stacked into $\boldsymbol{u}^\star\in\mathbb{R}^N$, and the LF map $\boldsymbol{\ell}\in\mathbb{R}^N$ follows the same ordering. Let $\Omega\subseteq\mathcal{V}$ contain all dense measurement locations. A subset $\Omega_m=\{i_1,\ldots,i_m\}\subseteq\Omega$ is randomly selected to emulate a sampling rate of $\rho=m/|\Omega|$. The observations in $\Omega_m$ satisfy
\begin{equation}
	\boldsymbol{y}=\boldsymbol{S}\boldsymbol{u}^\star+\boldsymbol{\epsilon},
\end{equation}
where $\boldsymbol{S}\in\mathbb{R}^{m\times N}$ is the sampling matrix with $\boldsymbol{S}_{j,:}=\boldsymbol{e}_{i_j}^\top$, $\boldsymbol{y}\in\mathbb{R}^m$ contains the observed power values, and $\boldsymbol{\epsilon}\in\mathbb{R}^m$ represents observation errors.

\subsection{Simulation-Based Background Space}\label{subsec:background_space}
The background space represents a family of fields informed by the LF simulation. Using the simulated map and a constant vector as basis functions, we define
\begin{equation}
	\mathcal{Z}=range(\boldsymbol{F}),\quad \boldsymbol{F}=\begin{bmatrix}
		\boldsymbol{\ell} & \boldsymbol{1}
	\end{bmatrix}\in\mathbb{R}^{N\times 2}.
\end{equation}
Each background field takes the form $\boldsymbol{z}=\boldsymbol{F}\boldsymbol{\beta}=a\boldsymbol{\ell}+b\boldsymbol{1}$, where $\boldsymbol{\beta}=\left[a,b\right]^\top$ contains the scale and offset coefficients. For a nonconstant LF map, $\mathcal{Z}$ is a two-dimensional linear space that retains the simulated spatial pattern while allowing global scale and offset adjustments. To accommodate discrepancies beyond these adjustments, the reconstructed field is expressed as
\begin{equation}
	\boldsymbol{u}=\boldsymbol{z}+\boldsymbol{\eta},\quad \boldsymbol{z}\in\mathcal{Z},
\end{equation}
where $\boldsymbol{\eta}\in\mathbb{R}^N$ is a spatial correction sought in the update space constructed in Sec.~\ref{subsec:update_space}.

\subsection{Graph-Based Update Space}\label{subsec:update_space}
Sparse observations alone do not uniquely determine the spatial update over the reconstruction grid. We therefore introduce a graph-based kernel to encode spatial regularity and construct the update space. An undirected weighted graph $\mathcal{G}=\left(\mathcal{V},\mathcal{E},\boldsymbol{W}\right)$ is defined on the original tow-dimensional grid. Its nodes represent non-building cells, and its edge connect horizontally or vertically adjacent non-building cells. The symmetric weight matrix $\boldsymbol{W}\in\mathbb{R}^{N\times N}$ has entries
\begin{equation}
	{W}_{i,j}=\begin{cases}
		d^{-1}_{ij}, & (i,j)\in\mathcal{E},\\
		0, & \text{otherwise},
	\end{cases}
\end{equation}
where $d_{ij}$ is the distance between cell centers in grid units. The graph Laplacian is $\boldsymbol{L}=\boldsymbol{D}-\boldsymbol{W}$, with $\boldsymbol{D}=diag(\boldsymbol{W}\boldsymbol{1})$ containing the weighted node degrees. Let $\bar{\boldsymbol{L}}=\boldsymbol{L}/d_L$, where $d_L$ is the median positive eigenvalue of $\boldsymbol{L}$. The spatial kernel is defined as
\begin{equation}\label{eq_kernel}
	\boldsymbol{K}_0=(\kappa^2 \boldsymbol{I}+\bar{\boldsymbol{L}})^{-s},\quad \boldsymbol{K}=\frac{\boldsymbol{K}_0}{tr(\boldsymbol{K}_0)/N},
\end{equation}
where $\kappa>0$ controls the spectral scale and ensures positive definiteness, while $s>0$ controls the decay of kernel eigenvalues. The normalization in Eq.~\eqref{eq_kernel} gives $\boldsymbol{K}$ a unit average diagonal value.

The kernel equips $\mathbb{R}^N$ with the inner product $\langle \boldsymbol{p},\boldsymbol{q}\rangle_K$. For the point-observation functional $\mathcal{L}_j(\boldsymbol{u})={u}_{i_j}$ at sampling location $i_j$, its Riesz representer is $\boldsymbol{q}_j=\boldsymbol{K}\boldsymbol{e}_{i_j}$, satisfying $\mathcal{L}_j(\boldsymbol{u})=\langle \boldsymbol{q}_j,\boldsymbol{u} \rangle_K$. These representers span the update space
\begin{equation}
	\mathcal{U}_{\Omega_m}=span\{\boldsymbol{q}_1,\ldots,\boldsymbol{q}_m\}=range(\boldsymbol{K}\boldsymbol{S}^\top).
\end{equation}
Thus, each update can be written as $\boldsymbol{\eta}=\boldsymbol{K}\boldsymbol{S}^\top\boldsymbol{\alpha}$, where $\boldsymbol{\alpha}\in\mathbb{R}^{m}$ contains coefficients estimated from the observations. The sampling locations determine the available kernel responses, while the kernel decides their spatial structure.

\subsection{Joint Optimization}
The background coefficients and spatial update are estimated jointly:
\begin{equation}\label{eq_joint_problem}
	(\widehat{\boldsymbol{\beta}},\widehat{\boldsymbol{\eta}}) = \underset{\boldsymbol{\beta}\in\mathbb{R}^2,\boldsymbol{\eta}\in\mathcal{U}_{\Omega_m}}{\arg\min}\;\left\{\frac{1}{2\nu}\|\boldsymbol{S}\boldsymbol{F}\boldsymbol{\beta}+\boldsymbol{S}\boldsymbol{\eta}-\boldsymbol{y}\|^2_2 + \frac{1}{2}\|\boldsymbol{\eta}\|^2_K\right\},
\end{equation} 
where $\nu>0$ balances observation fitting and update regularity. The joint problem is solved by first expressing the optimal update in terms of the background coefficients and then eliminating it.

Let $\boldsymbol{X}=\boldsymbol{S}\boldsymbol{F}$ and $\boldsymbol{C}_\nu=\boldsymbol{S}\boldsymbol{K}\boldsymbol{S}^\top+\nu\boldsymbol{I}$. For fixed $\boldsymbol{\beta}$, minimization over the update space gives
\begin{equation}
	\boldsymbol{\eta}(\boldsymbol{\beta})=\boldsymbol{K}\boldsymbol{S}^\top\boldsymbol{C}^{-1}_{\nu}(\boldsymbol{y}-\boldsymbol{X}\boldsymbol{\beta}).
\end{equation}
Substituting this expression into Eq.~\eqref{eq_joint_problem} yields
\begin{equation}\label{eq_beta_optimization}
	\widehat{\boldsymbol{\beta}}=\underset{\boldsymbol{\beta}\in\mathbb{R}^2}{\arg\min}\;\frac{1}{2}(\boldsymbol{y}-\boldsymbol{X}\boldsymbol{\beta})^\top\boldsymbol{C}^{-1}_{\nu}(\boldsymbol{y}-\boldsymbol{X}\boldsymbol{\beta}).
\end{equation}
Assuming $\boldsymbol{X}$ has full column rank, Eq.~\eqref{eq_beta_optimization} has the solution
\begin{equation}
	\widehat{\boldsymbol{\beta}}=(\boldsymbol{X}^\top\boldsymbol{C}_{\nu}^{-1}\boldsymbol{X})^{-1}\boldsymbol{X}^\top\boldsymbol{C}^{-1}_\nu\boldsymbol{y}.
\end{equation}
The spatial update is then revocer by evaluating $\boldsymbol{\eta}(\widehat{\boldsymbol{\beta}})$, giving the reconstructed field
\begin{equation}
	\widehat{\boldsymbol{u}}=\boldsymbol{F}\widehat{\boldsymbol{\beta}}+\boldsymbol{K}\boldsymbol{S}^\top\boldsymbol{C}^{-1}_{\nu}(\boldsymbol{y} - \boldsymbol{X}\widehat{\boldsymbol{\beta}}).
\end{equation}
The background estimate therefore depends on the kernel governing the update, preserving their coupling throughout the solution. The parameters $\kappa$, $s$, and $\nu$ are selected by cross-validation within $\Omega_m$, with both components refitted in each fold. Algorithm~\ref{alg:sapbdw} summarizes the whole procedure.

\begin{algorithm}[!t]
	\caption{SA-PBDW Radio Map Reconstruction}
	\label{alg:sapbdw}
	\begin{algorithmic}[1]
		\Require Building mask, LF map $\boldsymbol{\ell}$,
		sampling matrix $\boldsymbol{S}$, observations $\boldsymbol{y}$ with length of $m$,
		parameter candidates $\Theta$, and fold count $B$.
		\Ensure Reconstructed map $\widehat{\boldsymbol{u}}$
		
		\State Construct $\boldsymbol{F}$, the graph Laplacian and cache $\boldsymbol{K}$ for each candidate $(\kappa,s)$ using Eq.~\eqref{eq_kernel}.
		\State Partition observation indices $\{1,\ldots,m\}$ into folds $\{\mathcal{I}_b\}_{b=1}^{B}$.
		
		\For{each $\theta=(\kappa,s,\nu)\in\Theta$}
		\For{$b=1,\ldots,B$}
		\State Solve \eqref{eq_joint_problem} on observations outside $\mathcal{I}_b$, rebuilding the update space, to obtain $\widehat{\boldsymbol{u}}_b$.
		\State $E_b \gets \|\boldsymbol{S}_{\mathcal{I}_b,:} \widehat{\boldsymbol{u}}_b - \boldsymbol{y}_{\mathcal{I}_b}\|_2^2 / |\mathcal{I}_b|$.
		\EndFor
		\State $E(\theta)\gets\sqrt{B^{-1}\sum_{b=1}^{B}E_b}$.
		\EndFor
		
		\State Refit \eqref{eq_joint_problem} on all observations with $\widehat\theta=\arg\min_{\theta\in\Theta}E(\theta)$, to obtain $\widehat{\boldsymbol{\beta}}$ and $\widehat{\boldsymbol{\eta}}$.
		\State \Return $\widehat{\boldsymbol{u}} = \boldsymbol{F}\widehat{\boldsymbol{\beta}} + \widehat{\boldsymbol{\eta}}$.
	\end{algorithmic}
\end{algorithm}

\section{Experiment Results and Analysis}

\begin{figure}[!t]
	\begin{minipage}[b]{0.45\columnwidth}
		\centering
		\subfloat[]{
			\includegraphics[width=0.95\linewidth]{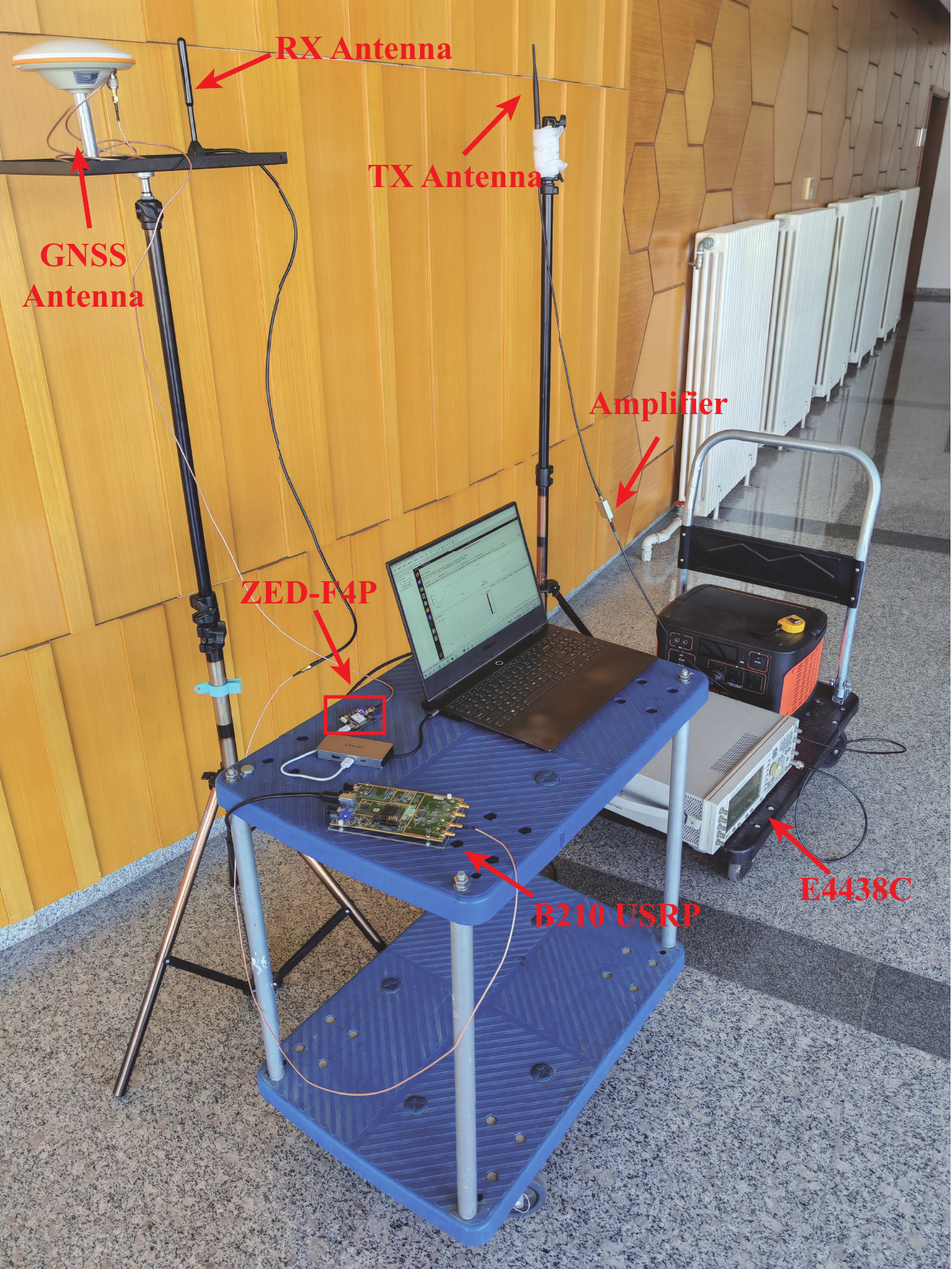}
			\label{fig_experiment_system}
		}		
	\end{minipage}
	\hfill
	\begin{minipage}[b]{0.55\columnwidth}
		\centering
		\subfloat[]{
			\includegraphics[width=\linewidth]{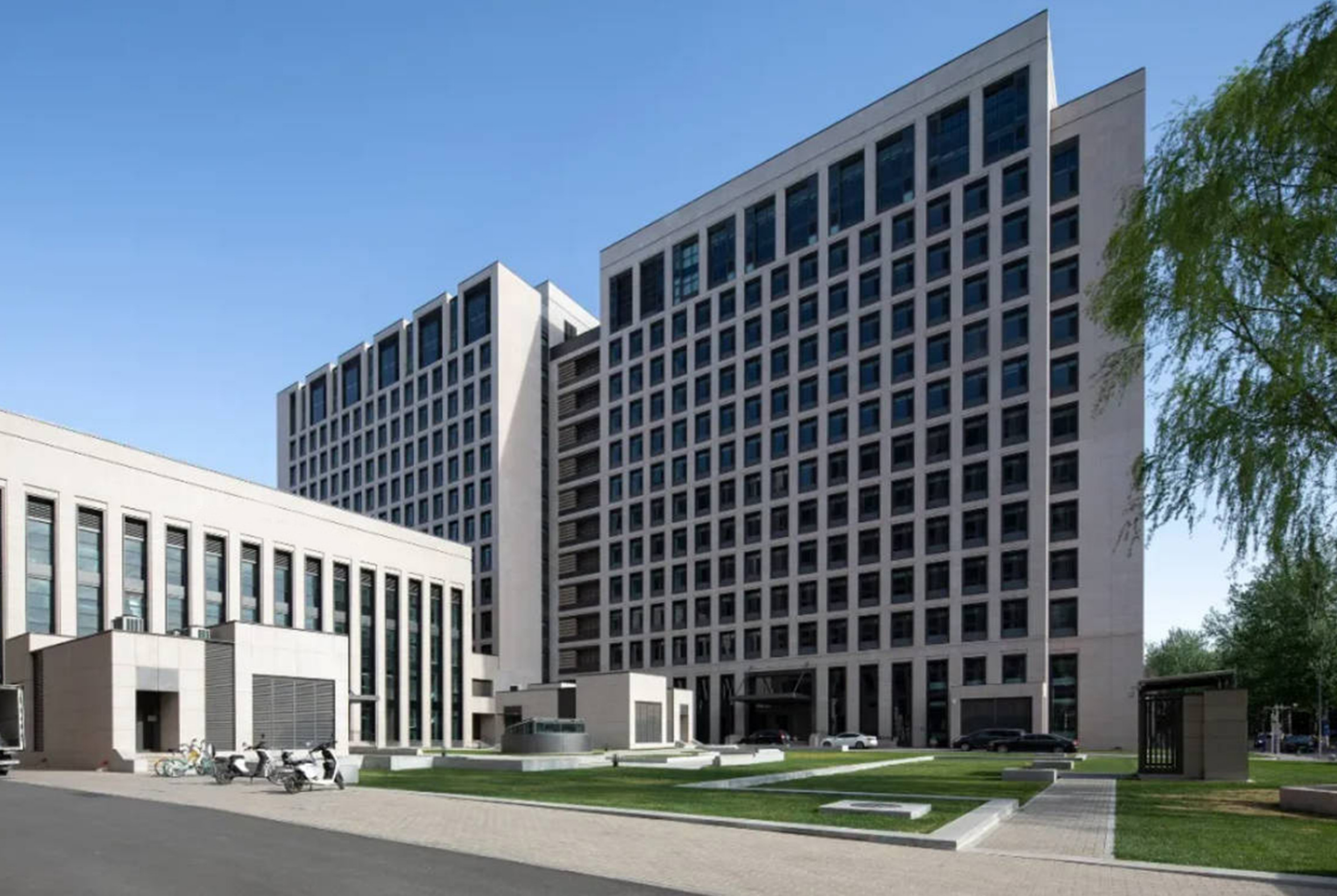}
			\label{fig_experiment_site}
		}
	\end{minipage}
	\caption{Field measurement system and site: (a) transmitter and receiver components; (b) overview of the experimental site.}\label{fig_experiment_overview}
\end{figure}
\begin{figure}[!t]
	\begin{minipage}[b]{0.49\columnwidth}
		\centering
		\subfloat[]{
			\includegraphics[width=\linewidth]{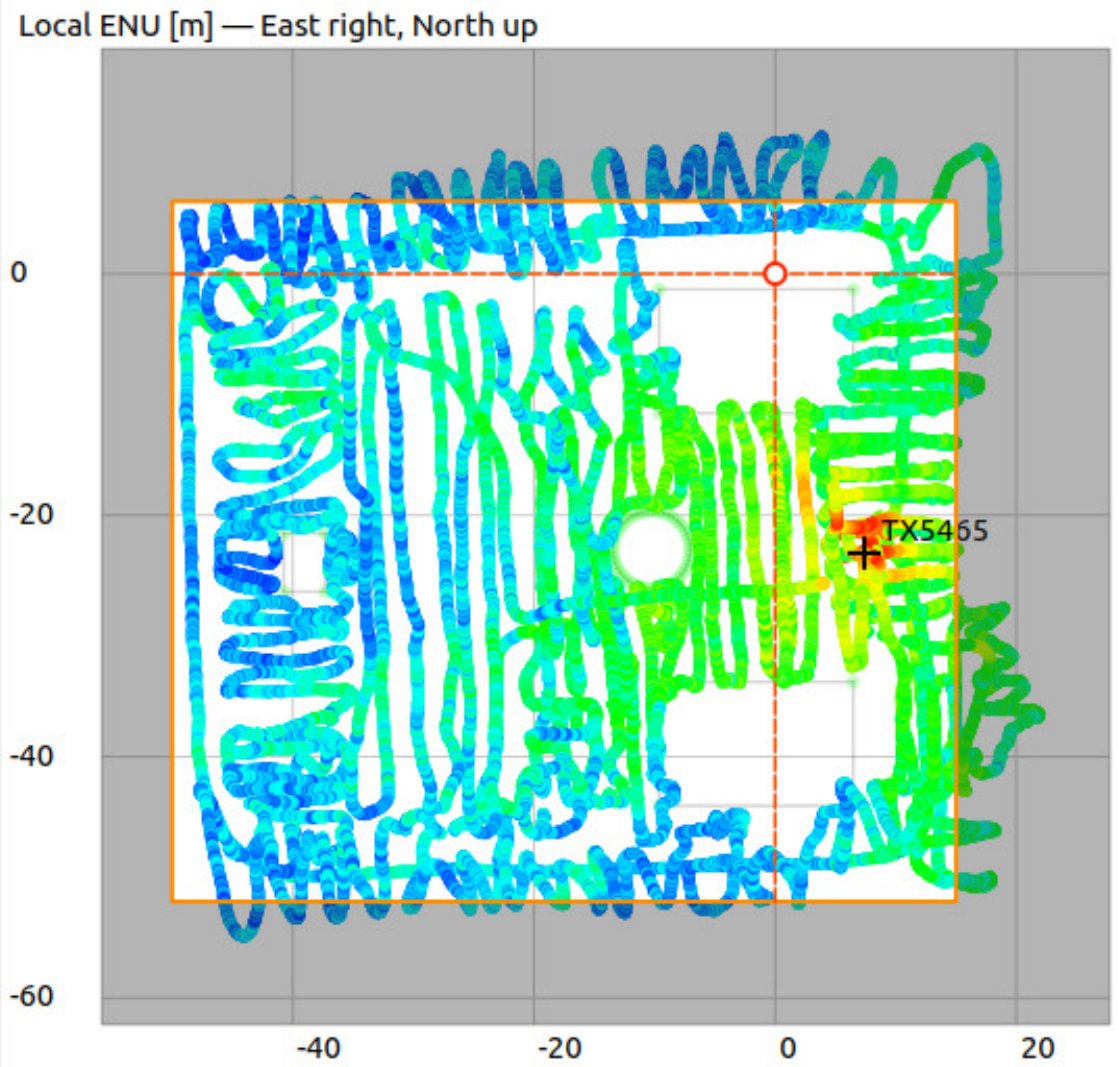}
			\label{fig_post_measurements}
		}		
	\end{minipage}
	\hfill
	\begin{minipage}[b]{0.49\columnwidth}
		\centering
		\subfloat[]{
			\includegraphics[width=\linewidth]{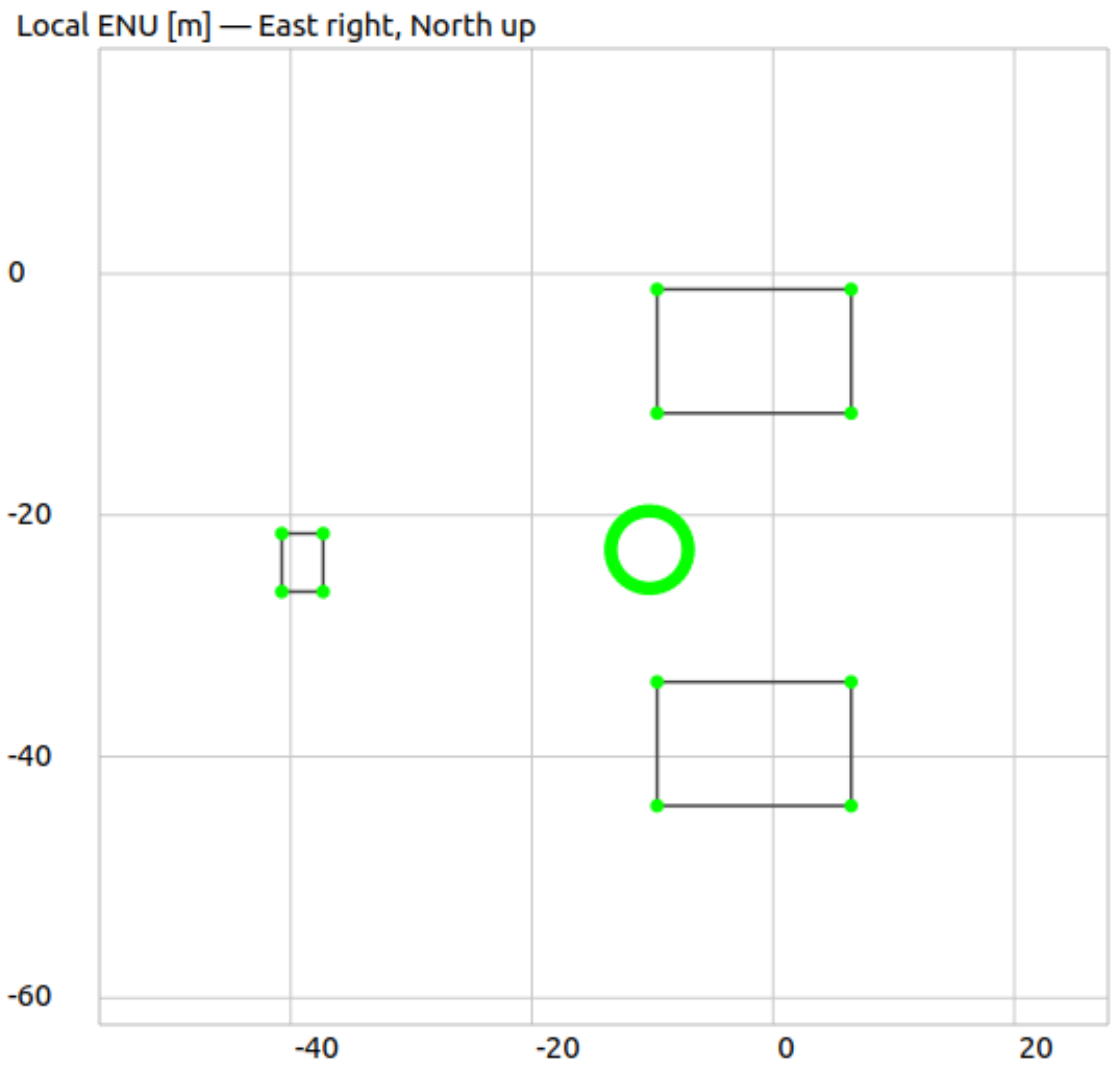}
			\label{fig_post_gnss}
		}
	\end{minipage}
	\caption{Measurement post-processing: (a) raw measurements and the selected mapping region; (b) manually refined building footprint coordinates. }\label{fig_post_software}
\end{figure}

\subsection{Datasets and Experimental Setup}
We first evaluate SA-PBDW on RadioMapSeer to assess its feasibility under controlled simulation conditions before using field measurements. We use DPM maps and ray tracing maps (referred to as IRT2) as LF inputs and HF references, respectively. Sparse observations are randomly sampled from non-building regions of the IRT2 maps. The baselines include classical IDW \cite{Denkovski2012idw} and Kriging \cite{sato2017kriging} methods, together with state-of-the-art Residual GPR (ResGPR) \cite{chen2025urban3dradiomap} and multi-fidelity GPR (MFGPR) \cite{cheng2027multifidelity} methods. Performance is evaluated using peak signal-to-noise ratio (PSNR), structural similarity index measure (SSIM), and root mean square error (RMSE) \cite{wang2024twostage} across different sampling rates.

For field evaluation, we construct an experimental system comprising a stationary transmitter and a mobile receiver unit to collect the FieldRadioMap, as shown in Fig.~\ref{fig_experiment_overview}\subref{fig_experiment_system}. Both omnidirectional antennas are maintained approximately \SI{1.8}{\metre} above ground. An Agilent E4438C signal generator continuously transmit a linear frequency-modulated (LFM) waveform centered at \SI{920}{\MHz} with a bandwidth of \SI{4}{\MHz} and a fixed output power of \SI{-17}{\dBm}. The receiver combines a USRP B210 for IQ acquisition with a real-time kinematic (RTK) positioning system comprising an HX-CSX627A global navigation satellite system (GNSS) antenna and a u-blox ZED-F9P module. Custom software running on Ubuntu associates the IQ data with positioning records. Received power of record $q$ is estimated using Welch's method with a Hamming window and a 4096-point fast Fourier transform (FFT):
\begin{equation}
	\begin{aligned}
		S_q[k]&=\frac{1}{Rf_sU_w}\sum_{r=1}^{R}\left|\sum_{n=0}^{L-1}w[n]x_{q,r}[n]\exp(-j\frac{2\pi kn}{N_{FFT}})\right|^2, \\
		P_q&=\frac{f_s}{N_{FFT}}\sum_{k\in\mathcal{K}}S_q[k],
	\end{aligned}
\end{equation}
where $N_{FFT}=4096$, $f_s$ is the sampling frequency, and $x_{q,r}[n]$ contains the IQ samples in the $r$-th of $R$ segments of length $L$ from record $q$. The window energy is $U_w=\sum_{n=0}^{L-1}|w[n]|^2$, and $\mathcal{K}$ contains the FFT bins corresponding to the target band. The experimental site is an open area, as shown in Fig.~\ref{fig_experiment_overview}\subref{fig_experiment_site}. A buffer zone seperates the measurement region from surrounding tall buildings to reduce reflection effects, while smaller buildings within the site provided a complex propagation environment.
\begin{figure*}[!t]
	\centering
	\includegraphics[width=\linewidth]{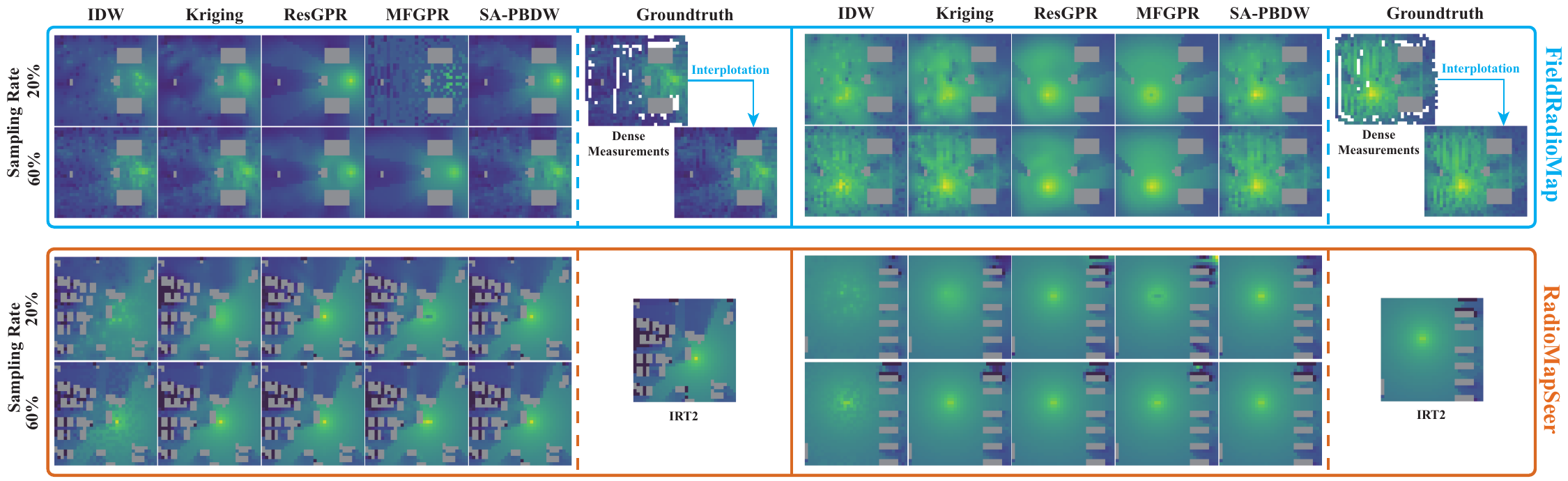}
	\caption{Reconstruction results at different sampling rates on FieldRadioMap and RadioMapSeer. White regions in the dense measurement maps of FieldRadioMap indicate missing measurements. These regions are interpolated only for visualization; evaluation excludes both unmeasured and building regions.}\label{fig_results}
\end{figure*}

\begin{table}[!t]
	\centering
	\caption{Field Measurement Configuration}
	\label{tab:field_config}
	\footnotesize
	\setlength{\tabcolsep}{3pt}
	\renewcommand{\arraystretch}{1.12}
	\begin{tabular}{llc}
		\toprule
		\textbf{Category} & \textbf{Parameter} & \textbf{Value/Description} \\
		\midrule
		\multirow[c]{6}{*}{Transmitter}
		& LFM bandwidth & \SI{4}{\MHz} \\
		& Carrier frequency & \SI{920}{\MHz} \\
		& Tx antenna type & Omnidirectional \\
		& Tx antenna height & \SI{1.8}{\metre} \\
		& Chirp duration & \SI{1}{\milli\second} \\
		& Generator output power & \SI{-17}{\dBm} \\
		\midrule
		\multirow[c]{4}{*}{Receiver}
		& Rx antenna type & Omnidirectional \\
		& Rx antenna height & \SI{1.8}{\metre} \\
		& GNSS antenna & Multi-band, multi-constellation \\
		& RTK positioning deviation & approx. 0.8--\SI{1.7}{\metre} (max. \SI{3.2}{\metre}) \\
		\midrule
		\multirow[c]{2}{*}{FieldRadioMap}
		& Measurement area & \SI{66}{\metre}$\times$\SI{58}{\metre} \\
		& Spatial resolution & \SI{2}{\metre} per pixel \\
		\bottomrule
	\end{tabular}
\end{table}
The collected discrete data are processed using custom software shown in Fig.~\ref{fig_post_software}\subref{fig_post_measurements}. To address severe position drift near buildings, building footprint coordinates are manually refined using repeated measurements to improve geometric consistency of the environment layout. Fig.~\ref{fig_post_software}\subref{fig_post_gnss} shows the corrected coordinates. The distinctive circular building and the symmetric square buildings on either side make it easy to match this layout to the real-world scene in Fig.~\ref{fig_experiment_overview}\subref{fig_experiment_site}. The measurement area is then partitioned into cells at a prescribed resolution, and the representative power of each occupied cell is calculated as
\begin{equation}
	\widehat{P}_i=\frac{1}{|\mathcal{A}_i|}\sum_{q\in\mathcal{A}_i}P_q,
\end{equation}
where $\mathcal{A}_i$ contains the measurement records assigned to cell $i$. These cell averages form the measured radio map, while occupied cells define the abovementioned $\Omega$. The grid resolution is chosen to balance spatial detail against positioning uncertainty. Detailed system parameters are provided in Table~\ref{tab:field_config}.

\begin{table}[!t]
	\centering
	\caption{Fidelity of DPM Maps on FieldRadioMap and RadioMapSeer}
	\label{table_dmp_evaluations}
	\begin{tabularx}{\linewidth}{lYYY}
		\toprule
		\textbf{Metrics} & \textbf{PSNR} & \textbf{RMSE} & \textbf{SSIM} \\
		\midrule
		FieldRadioMap & 21.12 & 0.091 & 0.49 \\
		RadioMapSeer & 23.28 & 0.071 & 0.77 \\
		\bottomrule
	\end{tabularx}
\end{table}

\begin{table*}[!t]
	\centering
	\caption{Evaluation metrics on FieldRadioMap and RadioMapSeer}
	\label{table_evaluations}
	\begin{tabularx}{\textwidth}{lcYYYYYYYYY}
		\toprule
		\multicolumn{2}{c}{\textbf{Sampling Rate}} & \multicolumn{3}{c}{20\%} & \multicolumn{3}{c}{40\%} & \multicolumn{3}{c}{60\%} \\ 
		\cmidrule(lr{0pt}){3-5}\cmidrule(lr{0pt}){6-8}\cmidrule(lr{0pt}){9-11}
		\multicolumn{2}{c}{\textbf{Metrics}} & \textbf{PSNR} & \textbf{RMSE} & \textbf{SSIM} & \textbf{PSNR} & \textbf{RMSE} & \textbf{SSIM} & \textbf{PSNR} & \textbf{RMSE} & \textbf{SSIM} \\ 
		\midrule
		\multirow{5}{*}{\textbf{FieldRadioMap}} & IDW & 23.91 & 0.065 & 0.59 & 25.64 & 0.053 & 0.71 & 27.59 & 0.042 & 0.81 \\
		& Kriging & 25.43 & 0.054 & 0.64 & 27.61 & 0.042 & 0.76 & 29.90 & 0.032 & \textbf{0.87} \\
		& ResGPR & 25.29 & 0.055 & 0.58 & 26.21 & 0.050 & 0.63 & 26.92 & 0.046 & 0.67 \\
		& MFGPR & 23.88 & 0.066 & 0.52 & 25.19 & 0.056 & 0.57 & 26.03 & 0.051 & 0.61  \\
		& SA-PBDW & \textbf{26.08} & \textbf{0.050} & \textbf{0.67} & \textbf{27.89} & \textbf{0.040} & \textbf{0.77} & \textbf{30.00} & \textbf{0.031} & \textbf{0.87} \\ 
		\midrule
		\multirow{5}{*}{\textbf{RadioMapSeer}} & IDW & 22.10 & 0.079 & 0.63 & 23.81 & 0.065 & 0.74 & 25.79 & 0.052 & 0.83 \\
		& Kriging & 23.29 & 0.069 & 0.72 & 25.64 & 0.053 & 0.84 & 28.13 & 0.040 & 0.91 \\
		& ResGPR & 25.78 & 0.052 & 0.84 & 27.83 & 0.041 & 0.90 & 29.99 & 0.032 & 0.94 \\
		& MFGPR & 23.37 & 0.069 & 0.72 & 24.79 & 0.058 & 0.79 & 26.26 & 0.049 & 0.84 \\
		& SA-PBDW & \textbf{26.30} & \textbf{0.049} & \textbf{0.86} & \textbf{28.10} & \textbf{0.039} & \textbf{0.91} & \textbf{30.34} & \textbf{0.031} & \textbf{0.95} \\ 
		\bottomrule
		\multicolumn{11}{l}{\textit{Notation}: The best results are highlighted in bold.}
	\end{tabularx}
\end{table*}

\subsection{Performance Analysis}
We evaluate SA-PBDW and all baselines on FieldRadioMap and RadioMapSeer through multiple independent trials. ResGPR, MFGPR and SA-PBDW use identical DPM maps as LF inputs on each dataset. Table~\ref{table_dmp_evaluations} summarizes the fidelity of these LF maps to the corresponding ground truth, establishing the reference performance before reconstruction. As shown in Table~\ref{table_evaluations}, incorporating sparse observations improves all three methods over the LF inputs. SA-PBDW achieves the highest PSNR and lowest RMSE at every tested sampling rate on both datasets. It also obtains the highest SSIM, sharing the best value with Kriging on FieldRadioMap at 60\% sampling. These results demonstrate consistent reconstruction performance across the scenarios and sampling rates.

Fig.~\ref{fig_results} provides representative visual comparisons. On FieldRadioMap, all methods generally recover the principal high-power regions and the overall spatial structure of the power distribution, but differ in their ability to reproduce local variations. ResGPR yields comparatively smooth distributions, while MFGPR exhibits localized artifacts at the lower sampling rate. SA-PBDW preserves the overall power distribution while more closely reproducing the spatial variations shown in the dense measurements. Its lower RMSE and higher SSIM support improvements in both numerical accuracy and structural fidelity. Kriging also captures these variations effectively as the second best approach on this dataset. In contrast, ground truth on RadioMapSeer exhibits smoother local distributions away from building-induced transitions. SA-PBDW and ResGPR perform best and second best, respectively, recovering the main spatial structures with high fidelity. 

What needs to be mentioned is that increasing the sampling rate improves reconstruction accuracy for all methods, with diminishing performance gaps at higher sampling rates. Therefore, the fidelity achieved by SA-PBDW under sparse sampling is particularly relevant to practical deployment, enabling reduced measurement requirements while preserving the characteristics of the target maps. This capability can facilitate field experiments where dense sampling is impractical and accelerate the development of measurement-based radio map datasets.

\section{Conclusion}
This letter proposed SA-PBDW for radio map reconstruction form propagation simulations and sparse field measurements. The non-neural method jointly estimates a simulation-based background and a spatial update. Dense field measurements were collected to construct FieldRadioMap dataset and provide spatially resolved references for evaluation. Experiments on FieldRadioMap and RadioMapSeer demonstrated consistent improvements over classical and state-of-the-art baselines, particularly in preserving local spatial variations under sparse sampling. These results support reduced field acquisition requirements and more efficient construction of measurement-based radio maps.

\bibliographystyle{IEEEtran}
\bibliography{references}{}

@ARTICLE{romero2022rme,
	author={Romero, Daniel and Kim, Seung-Jun},
	journal={IEEE Signal Process. Mag.}, 
	oldjournal={IEEE Signal Processing Magazine}, 
	title={Radio Map Estimation: A data-driven approach to spectrum cartography}, 
	year={2022},
	volume={39},
	number={6},
	pages={53-72},
	doi={10.1109/MSP.2022.3200175}
}

@ARTICLE{levie2021radiounet,
	author={Levie, Ron and Yapar, \c{C}a\c{g}kan and Kutyniok, Gitta and Caire, Giuseppe},
	journal={IEEE Trans. Wireless Commun.}, 
	oldjournal={IEEE Transactions on Wireless Communications}, 
	title={{RadioUNet}: Fast Radio Map Estimation With Convolutional Neural Networks}, 
	year={2021},
	volume={20},
	number={6},
	pages={4001-4015},
	doi={10.1109/TWC.2021.3054977}
}

@INPROCEEDINGS{Denkovski2012idw,
	author={Denkovski, D. and Atanasovski, V. and Gavrilovska, L. and Riihijärvi, J. and Mähönen, P.},
	booktitle={Proc. Int. ICST Conf. Cognitive Radio Oriented Wirel. Networks Commun., CROWNCOM},
	oldbooktitle={2012 7th International ICST Conference on Cognitive Radio Oriented Wireless Networks and Communications (CROWNCOM)}, 
	title={Reliability of a radio environment Map: Case of spatial interpolation techniques}, 
	year={2012},
	volume={},
	number={},
	pages={248-253},
	doi={10.4108/icst.crowncom.2012.248452}
}

@ARTICLE{sato2017kriging,
	author={Sato, Koya and Fujii, Takeo},
	journal={IEEE Trans. on Cogn. Commun. Netw.},
	oldjournal={IEEE Transactions on Cognitive Communications and Networking}, 
	title={Kriging-Based Interference Power Constraint: Integrated Design of the Radio Environment Map and Transmission Power}, 
	year={2017},
	volume={3},
	number={1},
	pages={13-25},
	doi={10.1109/TCCN.2017.2653189}
}

@ARTICLE{zhang2022gprradio,
	author={Zhang, Yifang and Wang, Shaowei},
	journal={IEEE Commun. Lett.}, 
	oldjournal={IEEE Communications Letters}, 
	title={K-Nearest Neighbors Gaussian Process Regression for Urban Radio Map Reconstruction}, 
	year={2022},
	volume={26},
	number={12},
	pages={3049-3053},
	doi={10.1109/LCOMM.2022.3207210}
}

@ARTICLE{zhang2020cbtd,
	author={Zhang, Guoyong and Fu, Xiao and Wang, Jun and Zhao, Xi-Le and Hong, Mingyi},
	journal={IEEE Trans. Signal Process.},
	oldjournal={IEEE Transactions on Signal Processing}, 
	title={Spectrum Cartography via Coupled Block-Term Tensor Decomposition}, 
	year={2020},
	volume={68},
	number={},
	pages={3660-3675},
	doi={10.1109/TSP.2020.2993530}
}

@ARTICLE{sun2026dun,
	author={Sun, Yazhou and Wang, Longhui and Chen, Xi and Wang, Jian and Shen, Yuan},
	journal={IEEE Trans. Netw. Sci. Eng.}, 
	oldjournal={IEEE Transactions on Network Science and Engineering},
	title={A Multi-Scale Dictionary Learning Deep Unfolding Network for Radio Map Reconstruction}, 
	year={2026},
	volume={13},
	number={},
	pages={10900-10917},
	doi={10.1109/TNSE.2026.3720605}
}

@ARTICLE{wang2024twostage,
	author={Wang, Yifan and Sun, Shu and Liu, Na and Xu, Lianming and Wang, Li},
	journal={IEEE Wireless Commun. Lett.},
	oldjournal={IEEE Wireless Communications Letters}, 
	title={Two-Stage Radio Map Construction With Real Environments and Sparse Measurements}, 
	year={2025},
	volume={14},
	number={4},
	pages={969-973},
	doi={10.1109/LWC.2025.3528512}
}

@ARTICLE{sun2025accuraterme,
	author={Sun, Yazhou and Wang, Longhui and Wang, Jian},
	journal={IEEE Sensors J.}, 
	oldjournal={IEEE Sensors Journal}, 
	title={An Accurate Radio Environment Map Reconstruction Method}, 
	year={2025},
	volume={25},
	number={12},
	pages={21988-22000},
	doi={10.1109/JSEN.2025.3562214}
}

@ARTICLE{chen2025urban3dradiomap,
	author={Chen, Xinwei and Zhong, Xiaofeng and Zhang, Zijian and Dai, Linglong and Zhou, Shidong},
	journal={IEEE Trans. Veh. Technol.},
	oldjournal={IEEE Transactions on Vehicular Technology}, 
	title={High-Efficiency Urban {3D} Radio Map Estimation Based on Sparse Measurements}, 
	year={2025},
	volume={74},
	number={10},
	pages={16488-16493},
	doi={10.1109/TVT.2025.3573595}
}

@ARTICLE{cheng2027multifidelity,
	author = {Cheng, Xuan and Wang, YaPeng and Yang, DingCheng and Ng, Benjamin K. and
	Im, Sio-Kei and Yang, Xu},
	title = {Multi-fidelity Gaussian process learning for 3D urban radio map
	reconstruction from sparse UAV measurements},
	journal={Ad Hoc Netw.},
	oldjournal = {AD HOC NETWORKS},
	year = {2027},
	volume = {194},
	month = {Jan},
	doi = {10.1016/j.adhoc.2026.104392},
}

@article{maday2015parameterized,
	title={A parameterized-background data-weak approach to variational data assimilation: formulation, analysis, and application to acoustics},
	author={Maday, Yvon and Patera, Anthony T and Penn, James D and Yano, Masayuki},
	journal={Int. J. Numer. Methods Eng.},
	oldjournal={International Journal for Numerical Methods in Engineering},
	volume={102},
	number={5},
	pages={933--965},
	year={2015},
	publisher={Wiley Online Library}
}

@ARTICLE{zhao2025radiodiff3d,
	author={Zhao, Le and Fei, Zesong and Wang, Xinyi and Luo, Jihao and Zheng, Zhong},
	journal={IEEE Wireless Commun. Lett.},
	oldjournal={IEEE Wireless Communications Letters},
	title={3D-RadioDiff: An Altitude-Conditioned Diffusion Model for 3D Radio Map Construction}, 
	year={2025},
	volume={14},
	number={7},
	pages={1969-1973},
	doi={10.1109/LWC.2025.3559934}
}

@INPROCEEDINGS{Shrestha2023rme,
	author={Shrestha, Raju and Ha, Tien Ngoc and Viet, Pham Q. and Romero, Daniel},
	oldbooktitle={2023 IEEE Conference on Antenna Measurements and Applications (CAMA)}, 
	booktitle={IEEE Conf. Antenna Meas. Appl. CAMA},
	title={Radio Map Estimation in the Real-World: Empirical Validation and Analysis}, 
	year={2023},
	volume={},
	number={},
	pages={169-174},
	doi={10.1109/CAMA57522.2023.10352759}
}

@inproceedings{wahl2005dominant,
	title={Dominant path prediction model for urban scenarios},
	author={Wahl, Ren{\'e} and W{\"o}lfle, Gerd and Wertz, Philipp and Wildbolz, Pascal and Landstorfer, Friedrich},
	booktitle={Proc. 14th IST mobile and wireless communications summit},
	address   = {Dresden, Germany},
	month = {June},
	pages={1--5},
	year={2005}
}

@ARTICLE{wang2025radiodiff,
	author={Wang, Xiucheng and Tao, Keda and Cheng, Nan and Yin, Zhisheng and Li, Zan and Zhang, Yuan and Shen, Xuemin},
	journal={IEEE Trans. Cogn. Commun. and Netw.},
	oldjournal={IEEE Transactions on Cognitive Communications and Networking}, 
	title={{RadioDiff}: An Effective Generative Diffusion Model for Sampling-Free Dynamic Radio Map Construction}, 
	year={2025},
	volume={11},
	number={2},
	pages={738-750},
	doi={10.1109/TCCN.2024.3504489}
}
	
\end{document}